\documentclass[a4paper,12pt]{article}    
\newcommand{\bm}[1]{\mbox{\boldmath   $#1$}}   
\usepackage{epsfig}
\begin{document}

\noindent {\bf X-ray Raman scattering study of aligned polyfluorene}\\

\noindent Szabolcs Galambosi$^{1*}$, Matti Knaapila$^2$, J.~Aleksi Soininen$^1$, Kim Nyg\aa rd$^1$, Simo Huotari$^3$, Frank Galbrecht$^4$, Ullrich Scherf$^4$, Andrew P.~Monkman$^2$, Keijo H\"am\"al\"ainen$^1$\\

\noindent $\mbox{}^1$ Division of X-ray Physics, Department of Physical
Sciences, University of Helsinki, POB-64, FI-00014 University of Helsinki,
Finland\\
$\mbox{}^2$ Department of Physics, University of Durham, South Road,
Durham DH1 3LE, United Kingdom\\
$\mbox{}^3$ European Synchrotron Research Facility, BP~220,
F-38043 Grenoble, France\\
$\mbox{}^4$ Fachbereich Chemie, Bergische Universit\"at Wuppertal, Gauss Strasse 20, D-42097 Wuppertal, Germany\vspace{1em}\\
{\small $\mbox{}^*$ email: szabolcs.galambosi@helsinki.fi\\
matti.knaapila@durham.ac.uk\\
aleksi.soininen@helsinki.fi\\
kim.nygard@helsinki.fi\\
simo.huotari@esrf.fr\\
frank@galbrecht.de\\
scherf@uni-wuppertal.de\\
a.p.monkman@durham.ac.uk\\
keijo.hamalainen@helsinki.fi}\\

\noindent{\bf Running title:} X-ray Raman scattering from polyfluorene\\

\section*{Abstract}

We present a non-resonant inelastic x-ray scattering study at the
carbon K-edge on aligned poly[9,9-bis(2-ethylhexyl)-fluorene-2,7-diyl]
and show that the x-ray Raman scattering technique can be used as a
practical alternative to x-ray absorption measurements. We demonstrate
that this novel method can be applied to studies on aligned $\pi$-conjugated
polymers complementing diffraction and optical studies. Combining the
experimental data and a very recently proposed theoretical scheme we
demonstrate a unique property of x-ray Raman scattering by performing
the symmetry decomposition on the density of unoccupied electronic
states into $s$- and $p$-type symmetry contributions.

\section*{Introduction}

Various synchrotron radiation based inelastic x-ray scattering methods
are now utilized in the study of electronic excitations. X-ray
Raman scattering (XRS) is a technique analogous to Raman scattering in
the optical region inasmuch as the energy difference of the incident
and the scattered photon is an indication of an excitation within the
target system. By using hard x rays an energy transfer of hundreds of
electron volts can easily be achieved leading to core electron
excitations in low-$Z$ elements.

The method differs from x-ray absorption spectroscopy (XAS), or more
specifically from near edge absorption fine structure spectroscopy
(NEXAFS) \cite{wende04} where the incident energy is tuned in the
vicinity of an absorption edge, for example, around 285~eV in the case
of carbon atoms. In an XRS experiment the energy of the incident
photon is typically around 10~keV, but the energy difference between
the incident and scattered photons is around 285~eV. The non-resonant
character of the core electron excitation leads to a relatively low
cross section, which necessitates the use synchrotron radiation to
perform XRS experiments. However, by using hard x rays, vacuum
sensitive samples can be measured as there is no need for vacuum
environment. Hard x rays are bulk sensitive, which in many cases
greatly simplifies the sample preparation and quality
requirements. XRS is analogous to inner-shell electron energy loss
spectroscopy (ISEELS) \cite{stohr92}. However, the use of photons as
the scattering probe instead of electrons leads to negligible multiple
scattering problems and again no vacuum environment is needed.

In an absorption process the energy loss and the photon momentum are
interconnected, but in XRS the momentum transfer during the scattering
process can be varied independently of the energy transfer simply by
varying the scattering angle. The cross section of XRS at low momentum
transfers can be related to the same dipole matrix elements which
determine the x-ray absorption process \cite{mizuno67}. This feature
has been successfully used in the field of hard condensed matter to
probe the electronic excitations in various low-$Z$ elements. XRS can be
used as a substitute for extended x-ray absorption fine structure
(EXAFS) \cite{tohji89, bowron2000}. In anisotropic materials, such as
graphite, analogously to polarization dependent x-ray absorption, XRS
may be used to probe the orientation of density of electronic states
\cite{nagasawa89}. The momentum transfer dependence of the XRS spectra
has, for example, been used to study excitations at the fluorine
K-edge in LiF \cite{galambosi02}, at the Be K-edge in beryllium
\cite{sternemann03} and recently at the B K-edge in the recently
discovered novel superconductor MgB$_2$ \cite{mattila2005}.

While organic materials being composed of light elements suit
inherently well to XRS studies, inelastic x-ray scattering has not
been utilized in this field to much extent. An important landmark was
achieved by Bergmann and co-workers by constructing an instrument with
sufficient energy resolution and a large angular acceptance of the
scattered photons \cite{bergmann01}. They have demonstrated that XRS
can be used as practical alternative to the XAS and the ISEELS methods
in the field of aromatic hydrocarbons \cite{gordon03, bergmann02}.
However, despite the advances in the XRS employed in the study of
small organic molecules, the XRS efforts in the research of
macromolecules, and electronically valuable $\pi$-conjugated polymers in
particular, seem to be scarce.
 
Polyfluorenes (PFs)\cite{scherf2002} are an important class of
$\pi$-conjugated polymers due to their efficient opto-electronic
performance and high stability.  Amongst PFs, branched side chain
poly[9,9-bis(2-ethylhexyl)-fluorene-2,7-diyl] (PF2/6) is an important
model compound as its photoluminescence and photoabsorption spectra
remain largely unchanged regardless of the long range order. Its
chemical structure is relatively simple and it is easily aligned to a
high degree \cite{knaapila2006}. This makes PF2/6 also technologically
interesting as sufficient alignment results in polarized
electroluminescence~\cite{grell1999} and enhanced charge carrier
mobility \cite{yasuda2005}. The self organized structure of aligned
PF2/6 has been comprehensively studied using electron diffraction
\cite{lieser2000}, fiber x-ray diffraction (XRD)\cite{knaapila2005b},
grazing-incidence x-ray diffraction (GIXRD) \cite{knaapila2005a} and
NEXAFS \cite{jung05} techniques.

In this paper we present an XRS study on aligned PF2/6. We have two
objectives; Firstly, we extend the use of XRS to $\pi$-conjugated
polymers in general and show that it is possible to make a distinction
between the contributions of backbone and side groups and connect this
to the macroscopic alignment of the polymer backbone complementing
diffraction and optical studies of aligned PF2/6.  Moreover, we show,
that in contrast to absorption methods, the momentum transfer
dependence of XRS can give truly unique information about the
electronic states of conjugated polymers. By combining the
experimental data to theoretical calculations according to a very
recently proposed theoretical scheme of Soininen {\em et al.}
\cite{soininen05}, we demonstrate that the unoccupied density of
electronic states can be decomposed into various symmetry components.

The energy units in this paper are in electronvolts and the momentum
units are in atomic units (1 a.u.~= $1/a_0$, where $a_0$ is the Bohr
radius).

\section*{Theoretical Section}
In a non-resonant inelastic x-ray scattering (NRIXS) process x rays
are scattered from the sample during which some amount of energy
$\omega$ and momentum ${\bf q}$ is transferred to the target system.
Within the non-relativistic Born approximation, the
double-differential cross section for this process is given by
\cite{schulke91}

\begin{equation}  \frac{{\mathrm   d}^2\sigma}{\mathrm  d  \Omega\,\mathrm   d\omega_2}=\left(\frac{\mathrm
d\sigma}{\mathrm  d\Omega}\right)_{Th}S({\bf q},\omega).
\label{crossection} 
\end{equation} 
The Thomson cross section $(\mathrm d\sigma/\mathrm d\Omega)_{Th}$
depends on the incident (scattered) photon energy $\omega_1$ ($\omega_2$) and polarization
vector $\bm\epsilon_1$ ( $\bm\epsilon_2$)

\begin{equation} \left(\frac{\mathrm d\sigma}{\mathrm
d\Omega}\right)_{Th} = r_0^2
(\bm\epsilon_1\cdot\bm\epsilon_2)^2\,\frac{\omega_2}{\omega_1},
\end{equation} 
where $r_0$ is the classical electron radius.  $S({\bf q},\omega)$ is
the dynamic structure factor that depends solely on the electronic
characteristics of the sample.

The non-resonant condition implies that the energy of the incident
photons is far from any absorption edges.  However, when the energy
{\em transfer} is close to the energy of an absorption edge (for
example 285~eV for the carbon K-edge), a spectral feature closely resembling an
absorption edge can be observed in the scattering spectra.  This
process is called x-ray Raman scattering.  The first
experimental reports \cite{dasgupta59,suzuki67} were followed by a
theoretical explanation \cite{mizuno67} and further experimental
studies \cite{tohji89,nagasawa89} that unambiguously proved that the XRS
spectrum measured at low momentum transfer values indeed gives
identical information compared to x-ray absorption spectra.

Within the quasiparticle approximation for excitations from a
tightly bound core state $|i\rangle$ the dynamic structure factor can
be written as \cite{schulke91} 

\begin{equation} S({\bf
q},\omega)=\sum_f |\langle f|e^{i{\bf q}\cdot{\bf
r}}|i\rangle|^2\,\delta(\omega+E_i-E_f), \label{sqomega}
\end{equation} 
where $|f\rangle$ is the photoelectron final state and $E_i$ ($E_f$)
is the initial (final) state energy.  For small momentum transfers,
the exponential in (\ref{sqomega}) can be expanded as

\begin{equation} e^{i{\bf q}\cdot{\bf r}}\approx 1 + i{\bf q}\cdot{\bf
r}+O(q^2).  \label{expapprox} \end{equation}
Due to the orthogonality of the initial and final states the first
term does not contribute to transitions and also the matrix elements
from the quadratic $q^2$ (and higher terms) are small.  Thus at low
momentum transfers, only the dipole allowed transitions contribute
significantly to the XRS spectrum.  Within this approximation the
similarity in the cross section of x-ray absorption
$\sigma(\omega_1)_{XAS}$ and $S({\bf q},\omega)$ of the scattering
process is evident

\begin{eqnarray} {\mathrm
\sigma}(\omega_1)_{XAS}\sim\omega_1\,|\bm\epsilon_1\cdot\langle f|{\bf
r}|i\rangle|^2\,\delta(\omega_1+E_i-E_f)\\\nonumber 
S({\bf q},\omega)=|{\bf
q}\cdot\langle f|{\bf
r}|i\rangle|^2\,\delta(\omega+E_i-E_f).\\\nonumber
\end{eqnarray}
As the momentum transfer is increased, the higher order terms in
Eq.~(\ref{expapprox}) no longer remain negligible.  Thus the spectral
features originating from non-dipole transitions gain weight, which
permits the study of electronic states not accessible with dipole
transitions.  

Theoretical methods for calculating the XRS spectra have been improved
tremendously in recent years.  Approaches like described in
Refs. \cite{soininen01a, soininen01b}, suitable for systems having a
periodic structure, have been successfully applied in the analysis of
XRS spectra from hard condensed matter systems \cite{galambosi02,
mattila05}.  For non-periodic soft condensed matter systems, such as
polymers, this approach is not applicable.  Real space multiple
scattering (RSMS) based methods such as the {\tt FEFF}-code
\cite{FEFF8} are widely used in the analysis of XAS experiments.  A
very recent report \cite{soininen05} showed how the momentum
dependence can be implemented into the RSMS framework, thus making it
possible to calculate the momentum transfer dependent XRS spectra for
non-periodic systems.

For core-electron excitations, the most important source of
error in the calculated XRS spectra is due to the inaccuracies in the
calculated density of states (DOS). On the other hand, the transition
matrix elements can be calculated to a relatively good accuracy.  To
address this problem, a new scheme was proposed by Soininen {\em et
al.}  \cite{soininen05}, in which the authors point out that by
exploiting the momentum transfer dependence of the experimental
spectra together with the calculated transition matrix elements, the
angular momentum projected DOS can be extracted from the data.  The method is
based on the fact that within a certain approximation the dynamic
structure factor can be written as

\begin{equation}  S({\bf  q},\omega)\sim\sum_l  c_l |M_l({\bf  q},\omega)|^2\rho_l(\omega),  
\label{symmetrydecompos}
\end{equation}
where $c_l$ is a coefficient, $M_l$ is the transition matrix element
and $\rho_l$ is the density of states having a symmetry
$l=s,p,d,\ldots$ This means that $S({\bf q},\omega)$ at each {\bf q},
can be regarded as the linear combination of DOS having different
symmetry, weighted by the relevant matrix elements.  With a suitable
inversion procedure the $\rho_l$ with different symmetries can be
found.  The current work relies on the fact that within the dipole limit
(at low $q$ values) only contribution to $S({\bf q},\omega)$ arises
from $p$-DOS ($p$-orbital contribution) which can be found from the
low momentum transfer spectra.  At higher $q$ values, other excitation
channels start to contribute.

In comparison with optical absorption spectroscopy the XRS method
provides a direct way to examine the unoccupied density of electronic
states. In XRS the initial $s$-type core state combined with the
dipole selection rule within the low $q$ regime result in transitions
purely into $p$-type unoccupied final states. In optical absorption
spectroscopy the dipole transitions also dominate, but the initial
electron state is different. The deep core electron states in
compounds retain their atomistic character but the valence states can
be heavily modified. In optical transitions the initial state is
formed by these valence electron states. Thus one usually cannot
expect a valence state to have a simple atomistic $s$- or $p$-type
symmetry, which renders the study of electronic states contributing to
the transitions somewhat difficult.  Moreover, the density of states
having other than a $p$-type symmetry can be extracted by employing
the momentum transfer dependence of XRS.

\section*{Experimental Section}  
\subsection*{Samples} 

\begin{figure}
\begin{center}
\epsfig{file=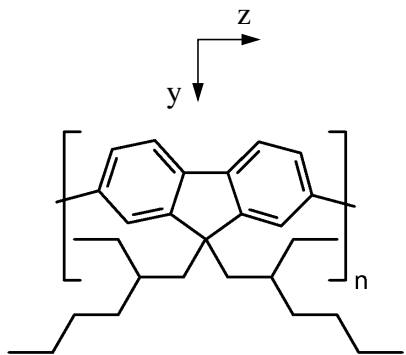,width=0.4\linewidth}
\caption{The chemical structure of
poly[9,9-bis(2-ethylhexyl)-fluorene-2,7-diyl] (PF2/6) along with the
coordinate axes used in the theoretical calculations. }
\label{fig1}
\end{center}
\end{figure}

PF2/6 ({$\it M_n$}=61 kg/mol, {$\it M_w$}=84 kg/mol) (Figure 1) was
prepared as described in Ref.  \cite{scherf2002}.  Polyfluorene fibers
were drawn above the glass transition temperature as shown in Ref.
\cite{knaapila2005a}.  The resulting rectangular samples had the
approximate dimensions of 30~mm$\times$7~mm$\times$1~mm
(length$\times$width$\times$thickness).  Each of the samples contained
a large number of fibers which, in an ideal case, would have the fiber
axes oriented along the longest side of the sample. The orientation
distribution of the fiber axes was determined using XRD by recording
the angular intensity distribution of the Debye rings using a
conventional x-ray tube pinhole camera and an image plate.

\subsection*{Measurements} 

\begin{figure}        
\begin{center}        
\epsfig{file=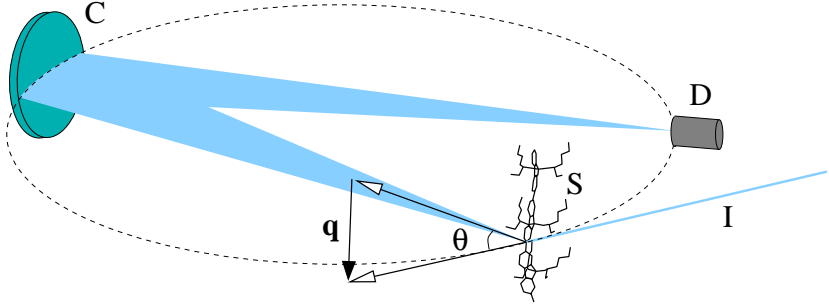,width=0.8\linewidth}        
\caption{The experimental setup. The incident x-ray beam (I) is
scattered by the sample (S). The analyzer crystal (C) focuses the
scattered photons having the correct energy into the detector
(D). Also shown is how the difference in the momenta of the incident
and the scattered photons defines the momentum transferred ({\bf q})
to the sample.}
\label{fig2}
\end{center} 
\end{figure}

\begin{figure}        
\begin{center}        
\epsfig{file=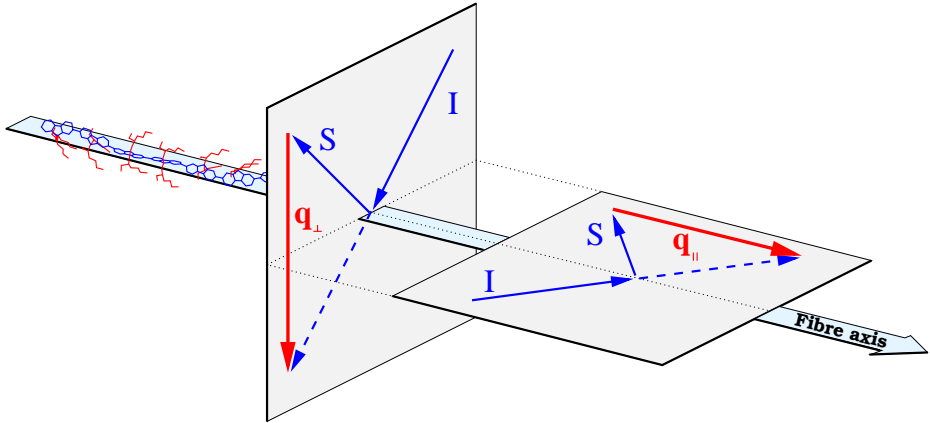,width=0.9\linewidth}        
\caption{A schematic representation of the two measurement geometries
used in the experiments. The direction of the incident (I) and
scattered (S) photons relative to the fibre axis can be chosen such
that the momentum transfer is either perpendicular  (${\bf q}_\perp$)
or parallel to (${\bf q}_{||}$) the fibre axis. }
\label{fig3}
\end{center}
\end{figure}

The inelastic x-ray scattering experiments were carried out on
beamline ID16 at the European Synchrotron Radiation Facility (ESRF),
Grenoble, France.  The x-rays produced by three consecutive undulators
were monochromatized using a double-crystal Si (111) monochromator.
The energy of the scattered photons was selected using a spherically
bent silicon crystal positioned on the Rowland circle (R=0.5~m)
utilizing the Si (555) reflection in a near backscattering
geometry. 
The experimental setup is shown schematically in Figure \ref{fig2}.
The energy loss spectra were recorded in the so-called inverse scan
technique \cite{hamalainen1996} in which the energy of the detected
photons ($\omega_2$) is kept fixed and the energy of the incident
photons ($\omega_1$) is varied.  A total energy resolution of 1.3~eV
was measured from the elastic line at 9885.6~eV.  
Two measurement geometries were used in which {\bf q} was perpendicular
 or parallel to the fibre axis (Figure \ref{fig3}). 
The momentum transfer resolution $\Delta q$ varied as a function of
the scattering angle and was 0.21~a.u.~at $q=1.4$~a.u.~and
0.06~a.u.~at $q=5.13$~a.u.~as calculated from the finite size of the
analyzer crystal which dominates the momentum resolution.

In order to reduce the possible degradation of the samples due to
radiation damage all x-ray scattering measurements were carried out in
vacuum and the samples were kept at a temperature of 20~K.  The size
of the x-ray beam at the sample was defocussed to 1.5 mm $\times$ 1.5
mm.  Several energy loss measurements each lasting about 20~min were
made from the same sample during which the consistency of the data was
continuously monitored. The total measurement time was about 6
hours/spectrum. The nature of radiation damage was studied further
{\it ex situ} by taking photoabsorption (UV-vis and Fourier-transform
infrared (FTIR)), photoluminescence (PL) spectra and XRD measurements of the
radiated spots after the synchrotron experiment.

The raw experimental spectra were corrected to account for the energy
dependence of the absorption occurring in the air paths, in the kapton
windows and in the sample.  The incident beam intensity was
continuously monitored to account for the slowly decaying synchrotron
ring current and for the energy dependence of the efficiency of the
optical elements. Each measured datapoint was normalized with respect
to the incident intensity and the dead-time of the detector was also
taken into account.  In addition to these standard corrections, the
background subtraction in our case needed special attention due to
other {\bf q}- and $\omega$-dependent electronic excitations
contributing to the inelastic x-ray spectrum.  In the low-$q$
measurements the carbon K-edge was well isolated from other
excitations leading to a small linear background.  For the large-$q$
spectra the K-edge, however, resides in the middle of the valence
scattering spectrum \cite{cooper04}, which renders the background
subtraction a non-trivial task.  To cope with this problem, we
performed {\em ab initio} Compton scattering calculations for one
monomer of PF using an extension \cite{hakala2004} of the {\tt
StoBe-DeMon} package \cite{stobe-demon}.  The results from the
theoretical calculations reproduced the experimental Compton profile
with a very good accuracy.  We then felt confident to use this
theoretical data to subtract the contributions of other than the
carbon 1s electrons from the experimental spectra.

\subsection*{Computational Section}

The theoretical calculations of the XRS spectra were performed using a
full multiple scattering formalism program based on the {\tt
FEFF8.2} package \cite{FEFF8}, modified to produce the momentum
transfer dependent x-ray Raman spectra \cite{soininen05}.  The
momentum transfer and direction dependent XRS calculations were
performed for carbon atoms in the central monomer in a PF polymer
consisting of three monomers.  The atomic coordinates were taken from
Ref.~\cite{knaapila2004}.

The schematic structure of PF2/6 and coordinate system used in the
calculations are shown in Figure \ref{fig1}. The coordinate
system was chosen such that $y$- and $z$-axes are in the plane of the
aromatic rings with the latter pointing approximately along the
polymer backbone. The $x$-axis is perpendicular to the planes of the
aromatic rings.

In the symmetry decomposition process the experimental data were first
brought to an absolute scale by normalizing them against the
calculated scattering cross section of an isolated atom
\cite{soininen05}. By assuming that purely dipole transitions are
allowed in the low-$q$ spectra, the $p$-type DOS was found by dividing
the low-$q$ spectra with the calculated matrix elements.  
The spectral contribution of transitions to $s$-type states was
extracted by subtracting the matrix element scaled $p$-DOS from the
high-$q$ spectra. The result was finally normalized using the proper
transition matrix elements to obtain the $s$-DOS.

\section*{Results and Discussion}

\begin{figure}        
\begin{center}        
\epsfig{file=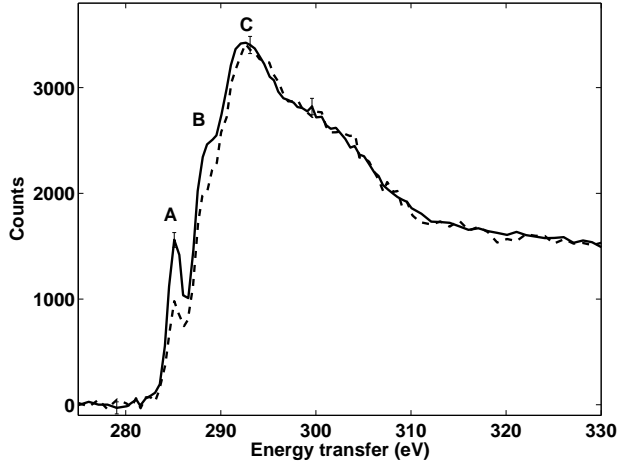,width=0.6\linewidth}        
\caption{The background subtracted experimental XRS spectra of
oriented PF 2/6 samples measured with {\bf q} perpendicular to (solid
line) and {\bf q} along (dashed line) the drawing axis of the
fibers. Selected errorbars due to counting statistics are shown. Peaks
A-C are described in the text.}
\label{fig4}
\end{center} 
\end{figure}

The experimental XRS spectra in the vicinity of the carbon K-edge at a
momentum transfer of 1.4~a.u. in two directions are shown in
Figure~\ref{fig4}. The results can be readily compared to x-ray
absorption spectra \cite{jung05}.  The two spectra are normalized to
the same area for energies over 295~eV. The only apparent difference
is the somewhat larger intensity in the 283-295~eV range when {\bf q}
is perpendicular to the drawing axis.  The experimental resolution is
high enough to discriminate several excitations.  Peak A at around
285~eV originates from the excitation of the carbon 1s electrons into
the $\pi^*$ orbitals of the aromatic rings in the backbone of PF
\cite{stohr92,jung05, agren95,dhez02,pattison2006}.  Feature B at
about 288~eV probably originates from excitations into $\sigma^*$
orbitals of the C-H bond.  Peak C residing around 293~eV is assigned
to transitions into $\sigma^*$ states on the C-C bond.

It is well established that transitions into $\pi^*$ states
responsible for excitations around 285~eV (peak A) are associated with
the presence of $\pi$ bonds \cite{dhez02}. Thus in the case of PF,
peak A can be uniquely identified to originate from the $\pi$ bonded
carbon atoms in the aromatic rings.  Other parts of the spectra are an
overlap of contributions from all carbon atoms.  As these $\pi^*$
orbitals are highly directional, the intensity variation as a function
of the scattering direction can be used to probe the orientation
distribution of the $\pi$-bonded molecules
\cite{pattison2006,stohr1987}.

The peak at 285~eV is visible in both measurement geometries
suggesting that the polyfluorene polymers have only partial
orientation preference along the drawing axis ($z$-axis). We estimated
the degree of orientation by performing x-ray diffraction study on the
angular intensity distribution of the (005) reflection
(cf. Refs.~\cite{knaapila2005b,knaapila2005a}). The intensity ratio of
this reflection along versus perpendicular to the drawing axis was
found to be 2.1:1, thus verifying that the orientation of the fibers
 was indeed only partial.

\begin{figure}
\begin{center}
\epsfig{file=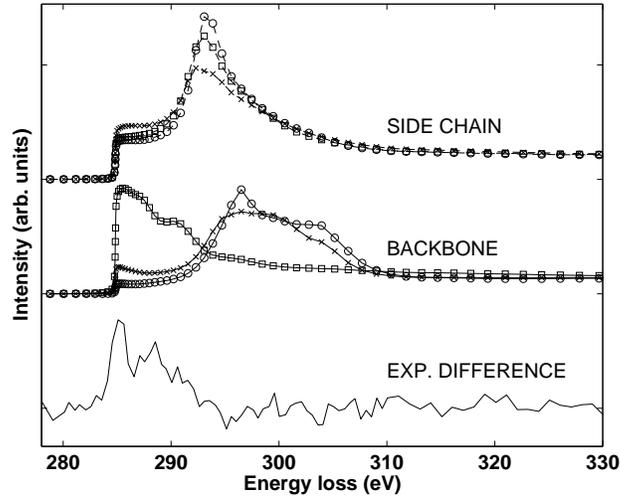,width=0.6\linewidth}
\caption{The calculated directional XRS spectra for a momentum
transfer of 1.4~a.u. The upper curves represent the calculated spectra
for the aliphatic sidechain carbon atoms and the middle ones for the
backbone carbon atoms with the momentum transfer aligned along the
$x$-axis (squares), the $y$-axis (crosses) and the $z$-axis
(circles). The depicted spectra represent the average over all
sidechain carbon atoms (dashed lines) and over all backbone carbons
(solid lines). The lower curve (thin solid line) shows the difference of
the experimental spectra in Figure \ref{fig4}.}
\label{fig5}
\end{center}
\end{figure}

To gain insight into the directional dependence of the XRS spectra, we
performed full multiple scattering calculations for all carbon atoms
in the middle monomer of a polyfluorene polymer consisting of three
monomer units. The calculated directional low-momentum transfer XRS
spectra averaged over the backbone carbon atoms and over the aliphatic
atoms are shown in Figure~\ref{fig5}. It is strikingly visible that
the directional dependence is predicted to be almost totally due to
the backbone carbons, namely the difference between the $yz$-plane and
the $x$-direction. In a polyfluorene polymer the consecutive monomers
are rotated relative to each other around the $z$-axis
\cite{knaapila2005b,knaapila2005a}. This means that even if we can
define the $z$-direction accurately, the physical properties
perpendicular to this axis can be regarded as an angular average of
the $xy$-plane. Thus by measuring the XRS spectra along the polymer
drawing axis one should probe mainly the electronic properties along
the $z$-direction. On the other hand, with {\bf q} perpendicular to
the drawing axis, one studies the properties averaged in the
$xy$-plane. We exploit the meridional orientation of our samples by
taking the difference of the spectra in Figure \ref{fig4}, i.e. the
spectra measured along the drawing axis and perpendicular to it.  As
shown in the lower part of Figure \ref{fig5}, the difference spectrum
is found to correlate very well with the theoretically calculated XRS
spectrum projected along the $x$-axis for the backbone atoms.  From
the calculated spectrum it is evident that for polyfluorene {\tt
FEFF8.2} does not reproduce the sharp peak at 285~eV. Thus, while
qualitatively it is clear that most of the direction dependence is due
to the difference between the $x$- and $z$-directions for the backbone
atoms, quantitative conclusions from the fine structures should not be
drawn.

\begin{figure}
\begin{center}
\epsfig{file=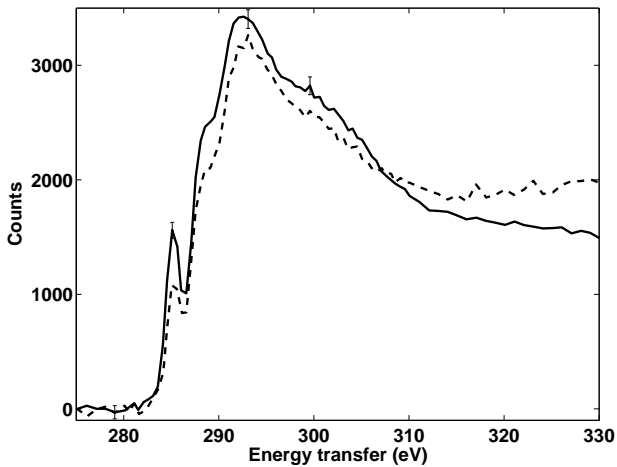,width=0.6\linewidth}
\caption{The XRS spectrum of PF with {\bf q} perpendicular to the
drawing axis at 1.4~a.u.~(solid line) and 5.13~a.u.~(dashed line).}
\label{fig6}
\end{center}
\end{figure}

The momentum transfer dependence of the XRS spectrum is shown in
Figure~\ref{fig6}.  The integrated intensity of the XRS spetrum grows
quickly with increasing $q$ \cite{schulke1988}. To emphasize the
differences of the spectra the data shown here are normalized to the
same area as is done in XAS \cite{wernet2004}.  The changes in the
spectrum are subtle, but we would like to point out that the
differences are due to the changes in the transition matrix
elements. The spectrum can no longer be regarded to originate purely
from dipole allowed transitions. From Figure~\ref{fig6} it is evident
that the higher energy regions gain considerable weight when compared
to the near edge region. Especially the relative intensity of the
285~eV peak is greatly diminished when compared to the maximum of the
spectrum.

\begin{figure}
\begin{center}
\epsfig{file=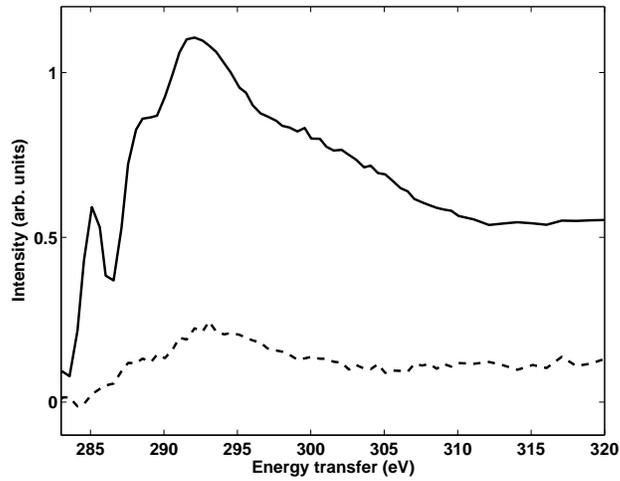,width=0.6\linewidth}
\caption{Final state angular momentum projected density of states of PF
2/6 showing the $s$-DOS (dashed line), $p$-DOS (solid line).}
\label{fig7}
\end{center}
\end{figure}

We carried out also XAS calculations using the {\tt FEFF8.2} program
\cite{FEFF8} and compared them to our calculated XRS spectra at
$q=1.4$~a.u. We found the two spectra virtually identical, confirming
that the low momentum transfer XRS spectra can be regarded to
originate from pure dipole allowed transitions from an $s$-type core
state to $p$-type valence states. Using the calculated matrix elements
together with the high momentum transfer spectrum, we can find the
$s$-type density of final states. Our main result is shown in Figure
\ref{fig7} along with the directional difference
spectrum. According to Figures~\ref{fig5} and \ref{fig7}, the peak at 285~eV
is due to electronic states that are perpendicular to the aromatic
rings having a pure $p$-type symmetry. These perpendicular states span
to about 10~eV over the edge with a slightly increasing $s$-symmetry
character. All other electronic states have, to some extent, an
$s$-type character along with the dominating $p$-type symmetry. The
largest contribution to the $s$-DOS happens around 293~eV coinciding
with the maxima in the calculated XRS spectra for the sidechain carbon
atoms. 

Now, compared to XANES, the extra information contributed by XRS is
evident. Not only can the equivalent of the absorption spectra
measured using hard x-rays, but the symmetry of the states leading to
the various spectral features can readily be found. In this context it
is intriguing to note that the pure $p$-type symmetry of the states
at 285~eV are in an excellent accord with what one would expect from
the simple H\"uckel molecular orbital model for the $\pi$ orbitals.

As described earlier in the experimental section, the measurements
were carried out in a vacuum environment at low temperatures with a
defocussed beam in order to reduce the radiation damage to the
samples. Despite these measures, after about 10~minutes of exposure a
slight darkening of the irradiated part of the sample could be
observed. As the data collection consisted of several 20-30 minutes
scans, we felt the need to investigate the possible degradation of the
samples.

\begin{figure}
\begin{center}
\epsfig{file=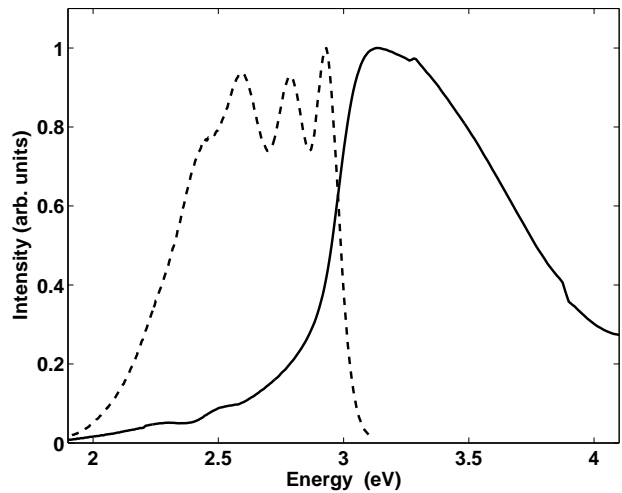,width=0.6\linewidth}
\caption{The photoluminescence (left) and absorption (right) spectra
from the irradiated part of PF 2/6. A small kink in the absorption
spectrum at 3.3~eV is an artifact form the spectrometer.}
\label{fig8}
\end{center}
\end{figure}

We performed several quick (2-3 minutes) XRS scan at the carbon K-edge
moving the footprint of the beam to a fresh spot for each scan. The
sum of the data from these scans were indistinguishable from the
actual data collection scans within counting statistics. We also
measured the valence excitation spectra both with shorter and longer
measurement times, but could not detect any time dependence of the
spectra in this case either. We thus concluded that either the
possible degradation is such that the current method is insensitive to
it or the sample degradation due to radiation damage occurs on a time
scale shorter than 2 minutes.

The intense x-ray beam is, however, expected to partially ionize the
polymer which may result in cross linking. This probably explains the
observed reduction in the solubility of the polymer which made the
molecular weight determination with gel permeability chromatography
(GPC) impractical.  A 10~nm red shift of the photoluminescence maximum
(from 370-385~nm to 380-395~nm) as well as an additional green
emission component beneath the original 0-0 and 0-1 peaks of the
polyfluorene PL spectrum were also observed. An example of these
spectra is shown in Figure \ref{fig8}. As the FTIR spectrum did not
indicate the presence of fluorenone groups, we expect that the
additional PL bands at 2.6/2.4 eV are predominantly not due to
well-known keto defects \cite{hintschich2003}. These bands are not
related to cross linking either, as simple cross linking of the side
chains will not result in colored products.  They may be an indication
of dibenzofulvene defects which are formed in a sequence of two
electron transfer and deprotonation steps in the absence of oxygen
\cite{nakano2001,monkman}.  It is yet noteworthy that the spectra
shown in Figure \ref{fig8} represent the worst case scenario after all
exposures and they are still dominated by the contribution of the
defect free polyfluorene.

X-ray diffraction spectra from the irradiated part of the samples
showed a somewhat decreased intensity of the diffraction maxima and
higher amorphous background when compared to the pristine sample
spots. However the angular distribution of the diffraction maxima were
unaltered.

From these studies we argue that the radiation damage does not impede
the successful interpretation of XRS data. The colorization of the
samples is not due to the well-known keto defects or simple cross
linking. It may arise from non-oxidatively formed dibenzofulvene
moieties. These defects apparently affect the crystallinity of the
sample but they do not alter the orientational distribution of the
polymers.

\section*{Conclusions}

  In summary, we have conducted an XRS study of aligned PF2/6 and
qualitatively interpreted the spectra based on the theoretical
framework.  In particular, we have shown that it is possible to make a
distinction between the contributions of backbone and side group
carbon atoms.  Moreover, we have shown, that the highly directional
nature of the electronic states of the aromatic carbon groups makes it
possible to study the orientation of $\pi$-conjugated polymers via XRS
thus complementing XAS, XRD and optical studies to some extent. We
have also demonstrated that the momentum transfer dependence of the
XRS spectrum can be utilized to find angular momentum projected
density of states.\\

{\bf Acknowledgments} This work has been supported by the Academy of
Finland (Contracts No. 201291/205967/110571) and One North-East (UIC Nanotechnology Grant).

%
%
%
%
\end{document}